\documentclass{article}
\usepackage[utf8]{inputenc}

\usepackage{xspace}
\usepackage{xcolor}
\usepackage{graphicx}
\usepackage{diagbox}
\graphicspath{ {./figures/} }
\usepackage{soul}

\title{Octopus: A Fair Packet Delivery Service}
\author{Junzhi Gong, Yuliang Li, Devdeep Ray, KK Yap, Nandita Dukkipati\\ \\Google LLC}
\date{}

\newcommand{\sys}{{Octopus}}
\newcommand{\mypara}[1]{\noindent{\bfseries {#1}.}}

\newcommand{\jg}[1]{#1}
\newcommand{\jgrm}[1]{}

\begin{document}

\maketitle

\section{Introduction}
\label{sec:introduction}
This paper focuses on \emph{multicast packet delivery fairness}.
There are several application, e.g., financial exchanges~\cite{cloudex,dbo}, consensus protocols~\cite{nezha}, and online-gaming~\cite{sync-ms}, where it is desirable for the multicast message to be delivered at the same time for all receivers.

Take the \jg{financial} exchange system as an example.
A \jg{financial} exchange system contains one centralized market engine, and a set of market participants.
The market engine is responsible to maintain stock information, process incoming orders from participants, and distributes market data periodically to participants.
Market participants subscribe to market data from the market engine, and submit stock orders to the engine on demand.
Participants usually compete to submit orders as fast as possible, due to the limited available stock share and its first-come-first-serve rule.
Therefore, it is critical for the market data to arrive at all subscribing participants at the same time, so that no participant has a structural advantage over other participants.  Such market requires a tight bound on the fairness, i.e., the packet arrival time between the first and last participants to receive it should be smaller than tens of nanoseconds.

To achieve such strict fairness guarantee, existing operators heavily customize their on-premise infrastructure.  Specifically, the network is connected via equal length cables to ensure all participants have the same network latency~\cite{costly-equal-cable}.  The network is also greatly over-provisioned to prevent any network congestion.  This results in a high-barrier of entry as the participants would need to build custom racks (which requires hardware expertise) and deploy within the on-premise location.

The challenges above motivates us to enable fair multicast packet delivery in an shared compute environment, e.g., in the Cloud, where the cost of business is significantly reduced.  This requires us to resolve any latency variations during the entire packet delivery process.
Foremost, latency variations can be introduced during the multicast itself.  When the multicast packet is delivered to different receivers, the sender might send the copies one at a time, resulting in each copy being released at different time.  Even with switch multicast support, variations can still exist.

Further latency variations are also introduced during packet forwarding, especially for a shared environment that is not greatly over provisioned, and interfering background traffic is more of the norm than exception.  In-network queueing and routing changes are all to be expected in such an environment, and both of them contribute to more latency variations.

There are several existing solutions attempting to mitigate the impact from packet delivery latency variations, but none of them provides a general fair packet delivery service with small delivery time variations.
CloudEx~\cite{cloudex} and Sync-MS~\cite{sync-ms} both provide a fair packet delivery service for any applications.
Between the sender and each receiver, they deploy one middlebox as an agent, to receive and hold multicast packets from the sender on behalf of receivers.
Then they have all agents release packets to receivers at the same time to ensure packet delivery fairness.
By holding and releasing packets at agents, they eliminate the impact from any latency variations during packet forwarding from the sender to agents.
However, they fail to provide good end-to-end packet delivery fairness, because they cannot control latency variations from agents to receivers.
More latency variations are introduced from packet processing inside agents, due to software performance jitters.

Other solutions like DBO~\cite{dbo}, Libra~\cite{libra}, and Frequent batch auctions~\cite{frequent-batch-auctions} attempts to redefine packet delivery fairness for specific applications (\textit{e.g.,} \jg{financial} exchange systems).
For example, Libra intentionally reshuffles the sequence of incoming multicast packets randomly for \jg{financial} exchange systems, so all receivers should suffer from a same level of packet delivery unfairness over a long time.
However, redefining fairness cannot work for all applications: reshuffling packet arrival sequence would hurt the performance of consensus protocols significantly.
Therefore, an ideal solution is to provide packet delivery services generally for all applications, without redefining the fairness.

Our goal is to propose a general but effective solution to provide packet delivery fairness in an shared compute environment, to release the high-barrier of entry from heavily customizing on-premise infrastructure.
We present \sys{}, a fair packet delivery framework that supports packet delivery fairness with delivery time variations smaller than tens of nanoseconds.
Similar to CloudEx and Sync-MS, \sys{} deploys agents between the sender and receivers, to hold and release packets to receivers at the same time.
Different from above existing solutions, one key insight of \sys{} is to repurpose a modern NIC capability, named time-based traffic shaping~\cite{intel-i210,intel-i225,nvidia-cx6,intel-i210-paper,carousel}, to release packets at predefined timestamp from the NIC hardware.
With the help of this hardware capability, \sys{} is able to eliminate latency variations introduced by software packet processing jitters at the agents.
Another key insight of \sys{} is to deploy agents on local SmartNICs at the receivers' servers, which reduces packet forwarding latency variations from agents to receivers.
With such design, applications can enjoy a fair packet delivery service in a shared compute environment.

We conduct evaluations for \sys{} in our seven-server testbed.
Our evaluation results show that with six receivers, \sys{} achieves good packet delivery fairness, with a 99.9\% tail maximum packet arrival time difference of around 30 nanosecond.
Such fairness is enough to support applications including \jg{financial} exchange systems.

\section{Problem Definition and Design Goals}
\label{sec:motivation}

In this section, we define the fair packet delivery problem in details and show a list of design goals for our fair packet delivery framework, \sys{}.

\noindent\textbf{Problem definition.}
For a fair packet delivery service, there are one packet sender and a set of $M$ packet receivers, each running on a separate process or a virtual machine.
There is no guarantee for the location running the sender and receivers.
When the packet sender multicasts a packet to all $M$ receivers, the fair packet delivery service should enable this packet to arrive at all receivers at the same time.

In this paper, we use the metric, \textit{packet delivery unfairness}, to evaluate fair packet delivery services.
To be specific, packet delivery unfairness is defined as the maximum packet arrival time difference for each multicast packet: given that the list of packet arrival time at receivers is $RT_1,RT_2,\cdots,RT_M$, the packet delivery unfairness is defined as
\[
max_{i=1}^{M}\{RT_i\}-min_{i=1}^{M}\{RT_i\}
\]
Ideally, such difference should be tens of nanoseconds to support applications with strict fairness requirements (like exchange systems).

\noindent\textbf{Design goals.}
We present the following design goals for our ideal fair packet delivery service.
\begin{enumerate}
    \item \textbf{Minimum packet delivery unfairness for packet arrival time.} This goal directly quantifies the fairness. To satisfy most of the applications, the ideal packet arrival time difference should be tens of nanosecond with a high probability (\textit{e.g.,} \jg{in our evaluation, \sys{} acheives less than 40 ns with a probability of} 99.9\%). 
    \item \textbf{Bounded packet latency.} It is critical to bound the packet latency between the sender to receivers, along with providing packet delivery fairness. 
    \item \textbf{Minimum application source code modification.} As a general service, it should provide an easy-to-use interface for applications without much modification to their source code. 
    \item \textbf{Minimum restrictions on application deployments.} We should not add many restrictions on deploying applications, including limiting locations of deployed applications and dedicated link paths.
\end{enumerate}
\section{\sys{} Design}
\label{sec:design}

In this section, we present \sys{}'s design details.
We first introduce \sys{}'s architecture and how a packet is multicasted to multiple receivers.
We then highlight \sys{}' key ideas, which address the state-of-the-art solutions' challenges.
Later, we present the details of our key ideas, and how they address the challenges.
Finally, we give a summary of \sys{}' advantages and potential limitations.

\subsection{System Architecture}

\begin{figure}
    \centering
    \includegraphics[width=\textwidth]{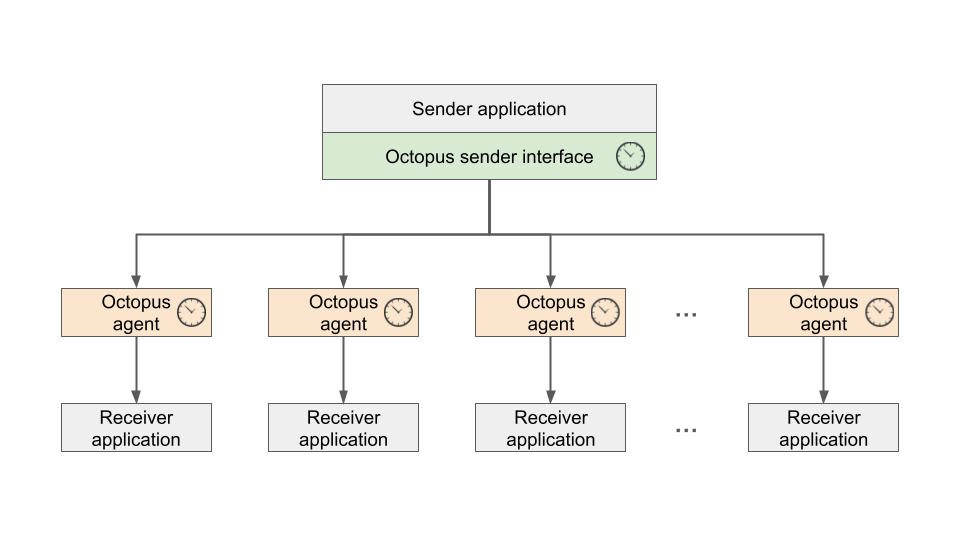}
    \caption{\sys{}' architecture. Note that grey boxes (the sender application and receiver applications) are not part of the \sys{}' architecture. Boxes marked with a clock are clock-synchronized.}
    \label{fig:architecture}
\end{figure}

Figure \ref{fig:architecture} shows a high-level architecture of \sys{}.
Given a sender application multicasting packets to multiple receiver applications, \sys{} consists of the following components: a sender interface used by the sender application, and one \sys{} agent for each receiver application.
\jg{Note that clocks are precisely synchronized for all components.}

\jgrm{Octopus first deploys a global clock synchronization service for the sender's server and all Octopus agents.
This synchronization service helps the sender and agents to agree on a same packet delivery time globally.}

\mypara{\sys{} agents}
Agents are responsible to control the packet release time towards receiver applications.
For each receiver application, one agent is deployed between the receiver and the sender, and it receives all incoming packets from the sender on behalf of its corresponding receiver.
Upon receiving each packet, the agent parses the intended release time provided by the sender (see below), and releases the packet on the intended time.

\mypara{\sys{} sender interface}
The sender interface is called by the sender application to start a fair packet delivery service.
For each packet to be multicasted, the sender interface is responsible to determine a same intended release time for all agents, and tags the timestamp along with the packet.

\mypara{Packet delivery steps}
In summary, \sys{} multicasts a packet in the following steps.
\begin{enumerate}
    \item The sender application generates a packet to multicast to all receiver applications, and calls the \sys{} sender interface to help multicast the packet.
    \item The \sys{} sender interface determines a global release time (which is calculated as the current timestamp plus a predefined delay $\delta$), and tags the timestamp along with the packet.
    \item The interface sends the tagged packet to all agents one by one.
    \item The agents receive the tagged packet, parse the release time, hold the packet, and release the packet to its corresponding receiver application at the release time.
\end{enumerate}

\subsection{Key Ideas}

The state-of-the-art solution, CloudEx, takes the first step to use agents sitting between the sender application and receivers to mitigate packet delivery unfairness in the cloud.
Similar to \sys{}' architecture, the key idea from CloudEx is to use agents to receive packets on behalf of receivers, hold those packets, and release those packets at a same predefined global timestamp.
As a result, latency variations between the sender and the agent does not introduce unfairness for packet delivery.

However, CloudEx still suffers from severe unfairness.
First, CloudEx's agents hold and release packets entirely in software, where large latency variations exist due to software performance jitters inside packet I/O libraries and drivers.
Such variations can be microsecond-level or even millisecond-level.
Additionally, latency variations also exist during packet forwarding from agents to receivers, potentially introduced by queuing delays from a set of switches.

Compared with CloudEx, there are two main key ideas for \sys{} to support a good fairness for the packet delivery service.

\begin{enumerate}
    \item \textbf{Time based packet releasing feature in NIC.} Many modern NICs support a hardware traffic shaping feature to control packet sending rate, which allows users to specify an intended release time for each packet. 
    This feature allows \sys{} to reduce latency variations introduced by software performance jitters at agents, contributing to higher packet delivery fairness.
    \item \textbf{Deploying agents on local SmartNICs.} \sys{} decides to deploy the agent on local SmartNICs at receivers' servers.
    Since the local SmartNIC is close to the receiver on the same server, latency variations are minimized between agents and receivers.
\end{enumerate}

\subsection{Accurate and Fair Packet Releasing with NIC Hardware Offloading}

In this section, we show how we repurpose a modern NIC capability to minimize latency variations introduced by software performance jitters at agents.

\mypara{Modern NICs' hardware capability}
NIC hardware capabilities are continuously evolving to meet new application needs and improve performance.
Traffic shaping is an important requirement for network applications.
To support high performance and flexibility, modern NIC vendors start to enable hardware traffic shaping capabilities for their NICs.
Specifically, many NIC vendors allows users to specify a release timestamp for each packet, and the NIC hardware holds the packet and releases it at the specified timestamp.
Applications can then implement any arbitrary traffic shaping pattern by defining the releasing time for each packet.

\mypara{\sys{}' design}
\sys{} minimizes latency variations at agents by repurposing the above traffic shaping capability.
Upon receiving a multicast packet from the sender, \sys{}' agent parses the intended release time from the packet.
Different from CloudEx which holds the packet in its software buffer, \sys{} directly sends the packet to the NIC hardware, tagged with the intended release time.

With this design, \sys{} eliminates most sources of latency variations from agent software, and simultaneously reduces overhead.
\sys{}' time-based packet releasing happens inside the NIC hardware, which eliminates any packet processing jitters from the software.
Additionally, \sys{}' agent becomes a stateless and light-weight application, reducing the probability of any performance jitters caused by state reads and writes.

\subsection{Minimize Unfairness after Packet Releasing}

After using the time-based packet releasing capability from the NIC hardware, \sys{} then needs to address latency variations after packets are released from agents.
CloudEx attempts to deploy its agents close to receivers (\textit{e.g.,} in the same pod), and has its agents to support multiple receivers.
However, as there is no guarantee on receivers' and agents' server locations, latency variations still exists at microsecond level due to in-network queuing delays and unequaled cable lengths.

Instead, \sys{} decides to run agents on the local SmartNIC with their corresponding receivers.
It allows the agent to either 1) deliver packets to the receiver on the host through ToR switch bounce back, or 2) deliver packets directly to the host using NICs' local forwarding feature.
As the agent on the local SmartNIC is close to the receiver, packets from the agent only travel through cables and the ToR switch under the same rack, where the latency variations are small 
\jg{(a few nanosecond variations without queueing)}.
For NICs supporting local forwarding, packets can be directly forwarded from the SmartNIC to the host, without going out via the ToR switch. 
In both cases, more sources of packet latency variations can be eliminated.

\subsection{Summary and Limitations}

\noindent\textbf{How \sys{} supports a good packet delivery fairness?}
With the above two insights, \sys{} eliminates most sources of packet latency variations compared with existing designs, under any type of networks \jg{(\textit{e.g.,} on-premise clusters, clouds)}.
First, with agents, latency variations from the sender to the agent does not contribute to any unfairness in the end-to-end packet delivery.
In addition, with the hardware traffic shaping capability, \sys{} eliminates latency variations introduced by software packet processing jitters at agents, where packets can be released at a predefined timestamp with a nanosecond-level accuracy.
Furthermore, after deploying agents at local SmartNICs with receivers, latency variations after packets are released from agents are also minimized.
To sum up, the end-to-end latency variations for multicast packets are minimized, contributing to a good packet delivery fairness.

\noindent\textbf{Tradeoff between packet latency and packet delivery fairness.}
\sys{}' sender interface determines the intended packet releasing time at agents by adding a predefined delay $\delta$ on the current timestamp.
The delay $\delta$ allows all copies of the packet to be received by all agents before releasing them, allowing all agents to release those copies at the correct time.
This means that a large $\delta$ provides better fairness guarantee, while increasing packet latency simultaneously.

\jg{Choosing an appropriate value for $\delta$ should rely on the end-to-end packet latency from the sender to agents. The first source of packet latency comes from software processing delay from the sender, allowing the software at the sender to push the packet to the NIC hardware. Additionally, after released by the sender, the packet forwarding delay to the agents is another source of latency, which includes path delay and queuing delay. Finally, the packet processing delay at agents should also be considered, so agents can successfully push the packet to the NIC hardware before the intended releasing time. To sum up, $\delta$ should be configured as the summation of the above three packet latencies, plus a small offset to allow more latency variations.}

Note that compared with heavily customized on-premise infrastructure with equal-length cables, \sys{} can easily guarantee a good fairness by setting $\delta$ according to the maximum packet processing delay inside agents plus a cable transmission delay.
Only if we want to share \sys{} with traffic from other applications, $\delta$ needs to be carefully chosen to tolerate the tail packet latency suffering from queuing delays.
In other words, \sys{} itself does not introduce the tradeoff, and the tradeoff only comes from the need to run applications in a shared environment like the cloud.

\noindent\textbf{Unfairness introduced by clock synchronization.} To let agents and the sender agree on the same global packet releasing time, their NIC hardware clocks need to be synchronized. Clock synchronization can never be perfect, and \sys{} also suffers from packet delivery unfairness introduced by synchronization inaccuracy. 
Fortunately, recent advances in clock sync has made nanosecond-level sync possible and practical \cite{sundial, graham, huygens}.

\mypara{\jg{Fairness impacted by the network load}}
\jg{The packet releasing time from \sys{} agents can be affected by the egress background traffic rate on the SmartNIC. The egress traffic can share the egress buffer with packets from \sys{} agents, resulting in additional queuing delay for after releasing. The impact of such egress traffic should be linear to the traffic rate, and we will leave the evaluation for such impact for future work.
If local forwarding is supported, the egress traffic should not affect fairness since the released packet does not goes though the egress buffer.}
\section{Evaluation}
\label{sec:evaluation}

\subsection{Evaluation Setup}

\mypara{Testbed}
We evaluate \sys{} in our cluster with a total of seven commodity servers.
Each server has one SmartNIC with its own CPU system that runs a Linux operating system, allowing users to deploy customized programs.

\mypara{Methods}
We write our own simple sender application to multicast packets to multiple receivers by using \sys{} sender interface.
We also write our own simple receiver application to receive those multicast packets from the sender application.
During the test, we deploy a sender application on one server, and deploy one receiver application on all other six servers.
For those six servers running receiver applications, we also deploy one \sys{} agent on its local SmartNIC.
All the above components run on a single CPU core, \jg{and there is no background traffic. Evaluations for the impact of background traffic is left for future work.}

For each test, we let the sender application to multicast a million packets to all six receivers via \sys{} agents.
To test how throughputs and $\delta$ affect \sys{}' packet delivery fairness, we configure \sys{}' throughput from 100 Kpps to 500 Kpps, and configure $\delta$ from 50 $\mu$s to 400 $\mu$s in different tests.
Note that for the throughput of 500 Kpps, the true sending rate of the sender application is 3 Mpps, since it needs to sends six copies for each multicast packet.

\mypara{Metric to evaluate}
To evaluate the fairness of \sys{}' packet delivery service, we collect the hardware arrival time for each multicast packet at all receivers, and calculate \textit{packet delivery unfairness} for each packet.

\subsection{End-to-end Results}

\begin{figure}
    \centering
    \includegraphics[width=0.7\textwidth]{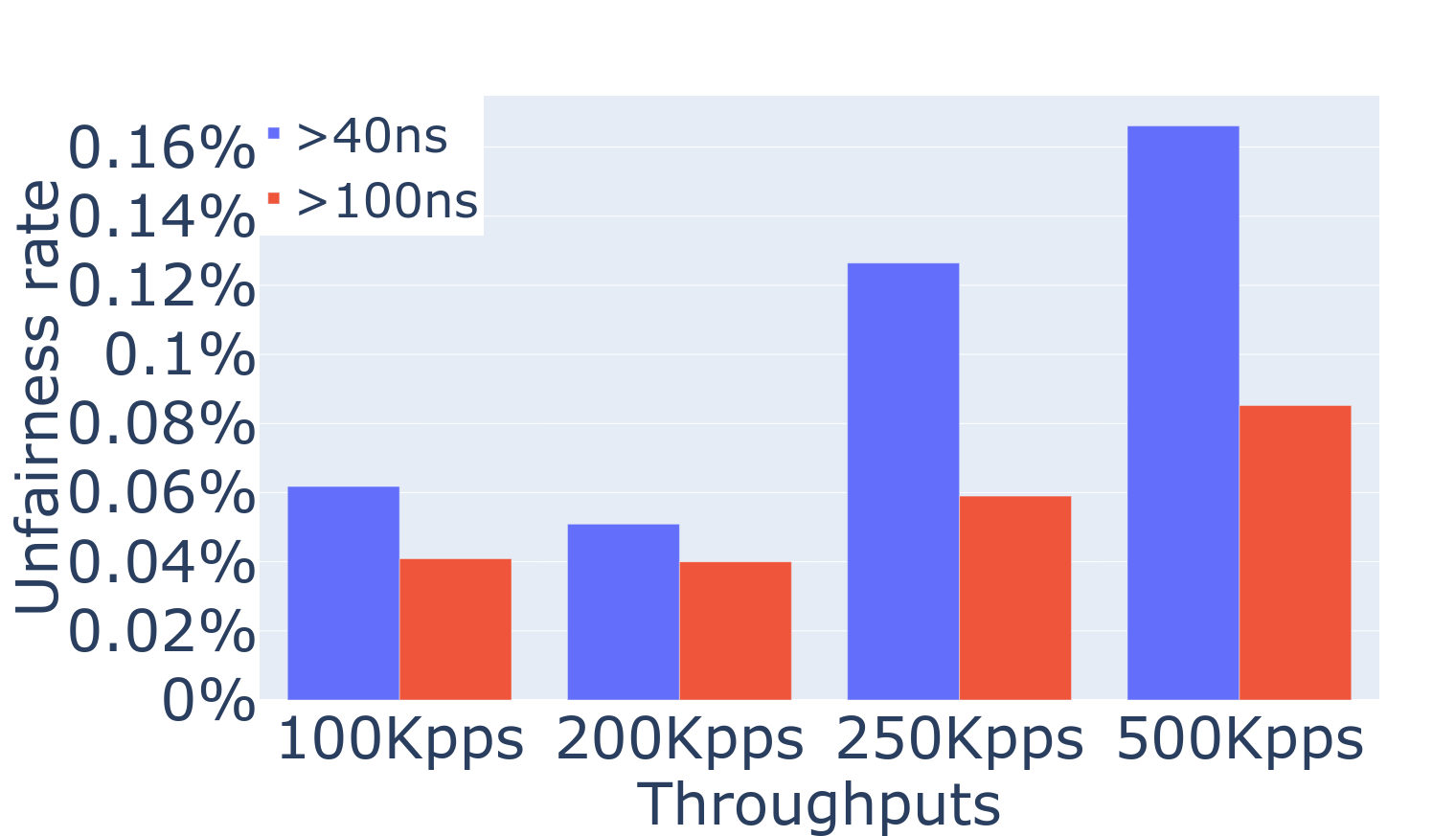}
    \caption{End to end evaluation results. The x-axis represents different throughputs to multicast packets at the sender, and the y-axis represents the unfairness rate ($>$40 ns and $>$100 ns) among receivers.}
    \label{fig:end_to_end}
\end{figure}

Figure \ref{fig:end_to_end} shows the end-to-end evaluation results for \sys{}.
In this test, we configure the \sys{}' sender application to multicast packets for a throughput from 100 Kpps to 500 Kpps, and set $\delta$ to 100 $\mu$s.
For the throughput of 500 Kpps, \sys{} only has 0.17\% packets whose unfairness is greater than 40 ns, and only has 0.09\% packets whose unfairness is greater than 100 ns.
For the throughput of 200 Kpps, \sys{} has only 0.05\% for unfairness rate ($>$40 ns), and 0.04\% for unfairness rate ($>$100 ns).

Compared with the state-of-the-art solution CloudEx, \sys{} successfully achieves a great packet delivery fairness by eliminating the packet forwarding jitters after released from agents.
\sys{} also takes advantages of the hardware traffic shaping feature to provide accurate packet releasing, without suffering from software performance jitters.

\begin{figure}
    \centering
    \includegraphics[width=0.7\textwidth]{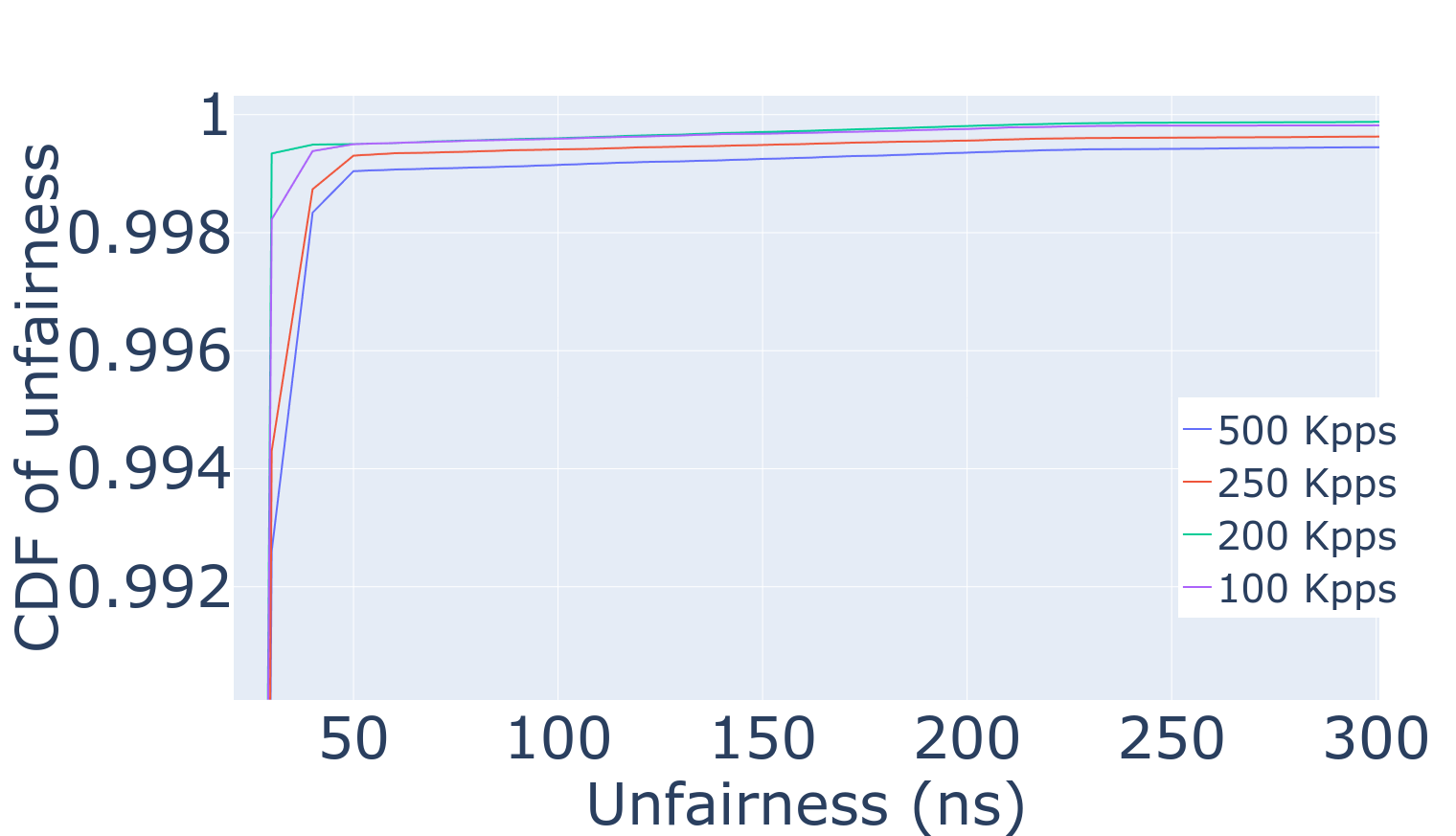}
    \caption{CDF of unfairness under different throughputs, with $\delta=100\mu$s.}
    \label{fig:cdf_100us}
\end{figure}

To better understand the good packet delivery fairness, we visualize the unfairness for each multicast packet in a CDF figure (see Figure~\ref{fig:cdf_100us}).
As shown in this figure, the packet delivery unfairness keeps between 0 to 30 ns for most of the packets, and only a very small portion of packets have unfairness greater than 50 ns.

\begin{table}
\begin{center}
\begin{tabular}{ c|c|c|c|c } 
 \hline
 \backslashbox{Throughput}{Tail unfairness (ns)} & 99\% & 99.9\% & 99.99\% \\
 \hline
 500 Kpps & 29 & 48 & 7852 \\ 
 250 Kpps & 27 & 40 & 612 \\ 
 200 Kpps & 25 & 29 & 397 \\
 100 Kpps & 25 & 35 & 987 \\
 \hline
\end{tabular}
\end{center}
\caption{Tail unfairness of \sys{} under different throughputs.}
 \label{tab:tail-unfairness}
\end{table}

Table~\ref{tab:tail-unfairness} shows the tail unfairness among all multicast packets under different throughputs, with $\delta$ equals to 100 $\mu$s.
Even under 500 Kpps, the 99\% and 99.9\% tail unfairness values are both under 50 ns.
The tail unfairness values for smaller throughputs are all between 25 ns and 40 ns.
For the 99.99\% tail unfairness, \sys{} has around 8 $\mu$s unfairness under 500 Kpps, and has an unfairness value under 1 $\mu$s when the throughput is smaller.
\jg{After careful inspection, the 99.99\% tail unfairness is introduced by late packet processing at the agent: due to the large packet rate and its large packet processing delay, the global time already passed the intended releasing time when some packets are processed by the agent.}

\subsection{The Impact from $\delta$}

\begin{figure}
    \centering
    \includegraphics[width=0.7\textwidth]{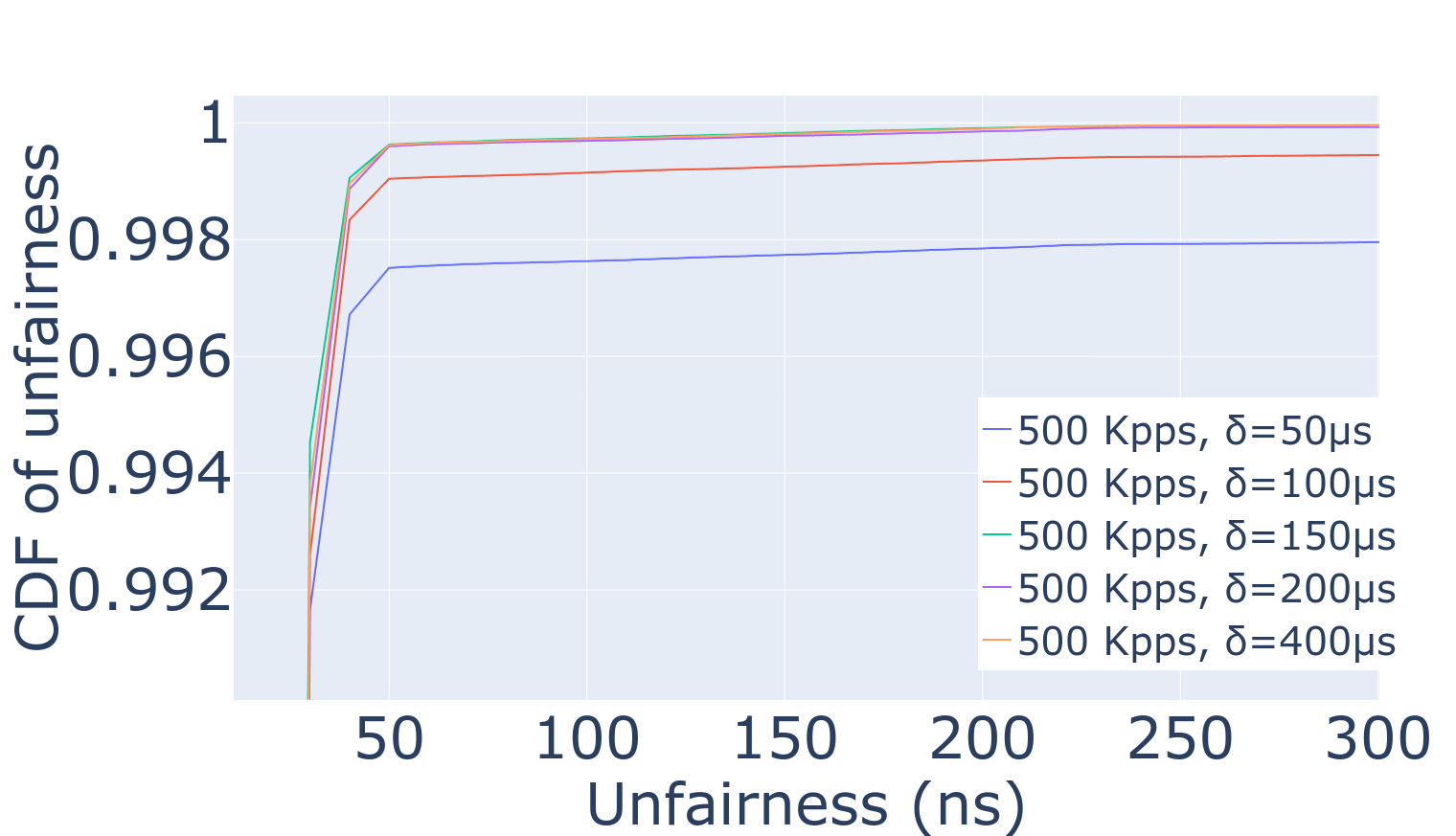}
    \caption{CDF of unfairness under 500 Kpps throughput.}
    \label{fig:cdf_500k}
\end{figure}

\begin{figure}
    \centering
    \includegraphics[width=0.7\textwidth]{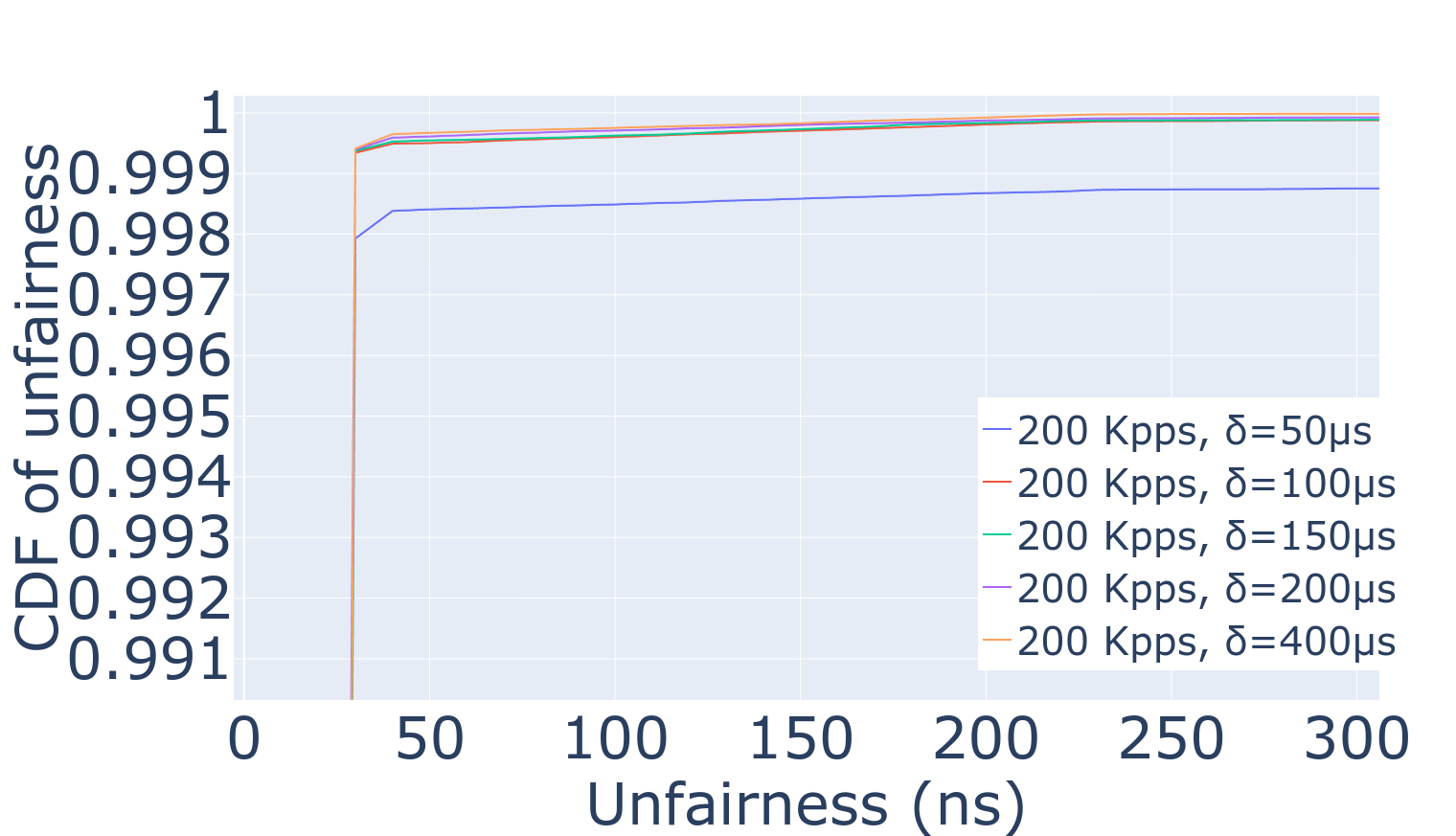}
    \caption{CDF of unfairness under 200 Kpps throughput.}
    \label{fig:cdf_200k}
\end{figure}

In this section, we evaluate \sys{} under different $\delta$ to show its impact.
Since our testbed is a shared environment, a small $\delta$ could potentially introduce packet delivery unfairness: a multicast packet arrive at an agent after the intended release time, and this agent can only release the packet after all other agents already release the packets to their receivers.
In this test, we choose five values for $\delta$, from 50 $\mu$s to 400 $\mu$s, and show the CDF of packet delivery unfairness among all multicast packets.

Figure \ref{fig:cdf_500k} and Figure \ref{fig:cdf_200k} show the CDF of unfairness under the throughput of 500 Kpps and 200 Kpps, respectively.
Under the throughput of 500 Kpps, \sys{} has an unfairness rate ($>$50 ns) smaller than 99.9\% only when $\delta$ is equal to 50 $\mu$s.
When $\delta$ is no less than 150 $\mu$s, the unfairness rate ($>$50 ns) keeps around 99.96\%.
Under the throughput of 200 Kpps, \sys{} has an unfairness rate ($>$50 ns) greater than 99.95\% when $\delta$ is no less than 100 $\mu$s.

In summary, when $\delta$ is no less than 100 $\mu$s, \sys{} achieves great packet delivery fairness with high probability.
Even if applications intends to reduce the latency to around 50 $\mu$s, \sys{} still achieves good packet delivery fairness for almost all multicast packets.

\section{Related Works}
\label{sec:related}

There are several existing solutions aiming to mitigating the impact from packet latency variations, but none of them can provide good packet delivery fairness for all applications.

\noindent\textbf{Solutions requiring deep software modifications.}
Some early solutions, like Libra~\cite{libra} and Frequent batch auctions~\cite{frequent-batch-auctions}, attempts to change the exchange software's internal logic to mitigate market data packet delivery fairness.
Libra proposes to randomize the processing sequence of orders already arrived to the market engine, reducing the probability of out-of-sequence orders from participants experiencing shorter network delay.
However, Libra's randomization only guarantees half of the orders to be processed in sequence.
Frequent batch auctions limits the market data multicast frequency to once every 100 ms, thus to mitigate the impact of varying packet delay among different forwarding paths.
However, this solution limits the performance of the exchange system.
Both solutions requires deep software modifications, and cannot generalize to all other applications.

\noindent\textbf{Solutions redefining packet delivery fairness.}
One of the state-of-the-art solutions, DBO~\cite{dbo}, mitigates packet delivery fairness for exchange systems, without the need for modifying the exchange software.
DBO measures the market data processing delay (from receiving the market data to submitting an order) at the vNIC of participants, and reorder different market orders based on their processing delays.
As a result, delivery time for market data does not affect the market fairness.
However, DBO requires to redefine the packet delivery fairness, and is specific to exchange systems.

\noindent\textbf{Solutions general to all applications.}
Another state-of-the-art solution, CloudEx~\cite{cloudex}, is general to all applications without modifying existing software.
Like \sys{}, CloudEx proposes to deploy agents to hold-and-release multicast packets for receivers.
The agent receives multicast packets on behalf of receivers, and synchronize the packet delivery towards receivers by releasing packets at the same time.
However, CloudEx cannot achieve good packet delivery fairness with packet arrival time variations smaller than tens of nanoseconds, due to large software performance jitters and nonidentical packet forwarding delays after released from agents.

\section{Conclusion}

The packet delivery fairness is critical in many applications in the cloud, such as exchange systems, consensus protocols, and online gaming applications.
However, due to nonidentical and dynamic packet forwarding paths, as well as many in-network queuing delays, supporting packet delivery fairness is challenging in a shared compute environment.
In this paper, we present \sys{}, the first general fair packet delivery service to achieve packet arrival time variations smaller than tens of nanoseconds, with the existence of latency variations in the network.
The key ideas of \sys{} to support such good fairness come from repurposing hardware traffic shaping capabilities in modern NICs, and deploying agents at local SmartNICs to minimize latency variations from packet forwarding.
Evaluation results show that \sys{} has less than 40 ns unfairness for up to 99.97\% multicast packets.

\clearpage
\bibliography{reference}

\begin{thebibliography}{10}

\bibitem{costly-equal-cable}
{Fast money: the battle against the high frequency traders}.
\newblock
  https://www.theguardian.com/business/2014/jun/07/inside-murky-world-high-frequency-trading.

\bibitem{intel-i210}
{Intel Ethernet Controller I210 Product Brief}.
\newblock
  https://www.mouser.com/pdfDocs/i210-ethernet-controller-family-brief.pdf.

\bibitem{intel-i225}
{Intel Ethernet Controller I225 Product Brief}.
\newblock
  https://cdrdv2.intel.com/v1/dl/getContent/621753?fileName=621753-Intel\%C2\%AE+Ethernet+Controller+I225-I226-Product+Brief-20230726.pdf.

\bibitem{nvidia-cx6}
{Nvidia Mellanox ConnectTx-6 Dx Datasheet}.
\newblock
  https://www.nvidia.com/content/dam/en-zz/Solutions/Data-Center/documents/nvidia-connectx-6-dx-en-hpe-datasheet.pdf.

\bibitem{intel-i210-paper}
Syed Sahal~Nazli Alhady and Amir Fuad~Wajdi Othman.
\newblock Time-aware traffic shaper using time-based packet scheduling on intel
  i210.
\newblock {\em International Journal of Research and Engineering},
  5(9):494--499, 2018.

\bibitem{frequent-batch-auctions}
Eric Budish, Peter Cramton, and John Shim.
\newblock The high-frequency trading arms race: Frequent batch auctions as a
  market design response.
\newblock {\em The Quarterly Journal of Economics}, 130(4):1547--1621, 2015.

\bibitem{nezha}
Jinkun Geng, Anirudh Sivaraman, Balaji Prabhakar, and Mendel Rosenblum.
\newblock Nezha: Deployable and high-performance consensus using synchronized
  clocks.
\newblock {\em arXiv preprint arXiv:2206.03285}, 2022.

\bibitem{huygens}
Yilong Geng, Shiyu Liu, Zi~Yin, Ashish Naik, Balaji Prabhakar, Mendel
  Rosenblum, and Amin Vahdat.
\newblock Exploiting a natural network effect for scalable, fine-grained clock
  synchronization.
\newblock In {\em 15th USENIX Symposium on Networked Systems Design and
  Implementation (NSDI 18)}, pages 81--94, Renton, WA, April 2018. USENIX
  Association.

\bibitem{cloudex}
Ahmad Ghalayini, Jinkun Geng, Vighnesh Sachidananda, Vinay Sriram, Yilong Geng,
  Balaji Prabhakar, Mendel Rosenblum, and Anirudh Sivaraman.
\newblock Cloudex: A fair-access financial exchange in the cloud.
\newblock In {\em Proceedings of the Workshop on Hot Topics in Operating
  Systems}, pages 96--103, 2021.

\bibitem{dbo}
Eashan Gupta, Prateesh Goyal, Ilias Marinos, Chenxingyu Zhao, Radhika Mittal,
  and Ranveer Chandra.
\newblock Dbo: Fairness for cloud-hosted financial exchanges.
\newblock In {\em Proceedings of the ACM SIGCOMM 2023 Conference}, pages
  550--563, 2023.

\bibitem{sundial}
Yuliang Li, Gautam Kumar, Hema Hariharan, Hassan Wassel, Peter Hochschild, Dave
  Platt, Simon Sabato, Minlan Yu, Nandita Dukkipati, Prashant Chandra, et~al.
\newblock Sundial: Fault-tolerant clock synchronization for datacenters.
\newblock In {\em 14th USENIX symposium on operating systems design and
  implementation (OSDI 20)}, pages 1171--1186, 2020.

\bibitem{sync-ms}
Yow-Jian Lin, Katherine Guo, and Sanjoy Paul.
\newblock Sync-ms: Synchronized messaging service for real-time multi-player
  distributed games.
\newblock In {\em 10th IEEE International Conference on Network Protocols,
  2002. Proceedings.}, pages 155--164. IEEE, 2002.

\bibitem{libra}
Vasilios Mavroudis and Hayden Melton.
\newblock Libra: Fair order-matching for electronic financial exchanges.
\newblock In {\em Proceedings of the 1st ACM Conference on Advances in
  Financial Technologies}, pages 156--168, 2019.

\bibitem{graham}
Ali Najafi and Michael Wei.
\newblock Graham: Synchronizing clocks by leveraging local clock properties.
\newblock In {\em 19th USENIX Symposium on Networked Systems Design and
  Implementation (NSDI 22)}, pages 453--466, Renton, WA, April 2022. USENIX
  Association.

\bibitem{carousel}
Ahmed Saeed, Nandita Dukkipati, Vytautas Valancius, Vinh The~Lam, Carlo
  Contavalli, and Amin Vahdat.
\newblock Carousel: Scalable traffic shaping at end hosts.
\newblock In {\em Proceedings of the Conference of the ACM Special Interest
  Group on Data Communication}, pages 404--417, 2017.

\end{thebibliography}
\bibliographystyle{plain}

\end{document}